\documentclass[preprint,showpacs,preprintnumbers,amsmath,amssymb]{revtex4}
\usepackage{graphicx}
\usepackage{dcolumn}
\usepackage{bm}
\usepackage{color}

\newcommand{\be}{\begin{eqnarray}}
\newcommand{\ee}{\end{eqnarray}}

\newcommand{\benl}{\begin{eqnarray*}}
\newcommand{\eenl}{\end{eqnarray*}}

\begin{document}
\title{Chiral Symmetry Breaking in Bent-Core Liquid Crystals}
\author{Lech Longa$^{1}$}
\hspace{-3cm}\email[e-mail address:]{lech.longa@.uj.edu.pl}
\author{Grzegorz Paj\c{a}k$^{1}$}
\author{Thomas Wydro$^{1,2,}$}
\email[e-mail address:]{wydro@univ-metz.fr}

\affiliation{$^{1}$Marian Smoluchowski Institute of Physics,
Department of Statistical Physics and  Mark Kac Center for Complex
Systems Research, Jagiellonian
University, Reymonta 4, Krak\'ow, Poland\\
$^{2}$Universit\'e Paul Verlaine -- Metz, Laboratoire de Physique
Mol\'{e}culaire et des Collisions,\\ 1 bvd Arago, 57078 Metz, France}
\date{\today}
\begin{abstract}
By molecular modeling we demonstrate  that the nematic long-range order discovered in bent-core liquid crystal systems should reveal further spatially homogeneous phases. Two of them are identified as a tetrahedratic nematic ($N_T$) phase  with $D_{2d}$ symmetry and a chiral tetrahedratic nematic  ($N_T^*$) phase with $D_2$ symmetry. These new phases were found for a lattice model with quadrupolar and octupolar anisotropic interactions using Mean Field theory and Monte Carlo simulations. The phase diagrams exhibit tetrahedratic ($T$), $N_T$ and $N_T^*$ phases, in addition to ordinary isotropic ($I$), uniaxial nematic ($N_U$) and biaxial nematic ($N_B$)  phases. To our knowledge, this is the first molecular model with  spontaneous chiral symmetry breaking in non-layered systems.
\end{abstract}
 \maketitle
%
%
%
%
%
The chirality of liquid crystals is commonly attributed to the presence of
optically active (chiral) molecules \cite{deGennes}.
Such \emph{intrinsic molecular chirality} distinguishes right-handed and left-handed molecules leading to a rich variety of chiral phases, where cholesteric phase \cite{deGennes} and  blue phases \cite{Englert} serve as typical examples.

Interestingly, this classical liquid crystalline chirality should be revised for \emph{achiral} bent-core (banana-shaped) liquid crystalline molecules \cite{Link,Walba,Selinger,Pelzl,Lubensky,Niori,Brand,Takezoe,Yan}. These fascinating compounds have substantially different physical properties than the molecules of the traditional calamitic materials, and have recently been a source of discoveries that are changing our view on molecular self-organization in liquid crystals \cite{Link,Walba,Pelzl,Niori,Takezoe}.

Several observations indicate that achiral bent-core molecules can acquire conformational chirality in layered (smectic)  mesophases \cite{Link,Takezoe}.
They can form ferroelectric and \emph{homogeneously} chiral states \cite{Walba}, or coexisting left and right-handed chiral domains in a nematic phase \cite{Pelzl,Niori}.
All basic theoretical interpretations of these observations  introduce at a macroscopic scale a tetrahedratic order parameter field. The interplay between this order parameter and the (nematic) quadrupolar tensor order parameter \cite{Longa1,Longa2} appears to be essential for the emergence of the chirality of the achiral bent-core molecules \cite{Lubensky,Brand}.
In the bent-core molecular systems this coupling generates not only the most interesting chiral symmetry breaking. The new tetrahedratic ($T$) and tetrahedratic nematic ($N_T$) phases should also be stabilized \cite{Lubensky}. Recently, the experimental indication of stable  $T$ and $N_T$ phases was indeed reported \cite{Wiant}, along with earlier discovery of the biaxial nematic phase \cite{Madsen,Acharya}.

This letter analyzes some consequences of molecular interactions that have tetrahedratic, uniaxial  and biaxial  components.
For a complete understanding of the macroscopic properties of mesophases in the bent-core systems, a first-rank tensor generating structures with polar order, is also required.
To simplify the analysis, the microscopic model being discussed  is limited to
quadrupolar  and octupolar interactions.
In addition to isotropic and nematic phases, the model is shown to stabilize $T$, $N_T$ and chiral tetrahedratic nematic ($N_T^*$) phases. These results are what we believe a first microscopic demonstration of   spontaneous chiral symmetry breaking in non-layered bent-core liquid crystals.

We assume the molecules to occupy the sites of a  three dimensional simple
cubic lattice with interactions limited to nearest neighbors (coordination number: $d=6$). The total interaction potential is then given by
   $ H = \frac{1}{2} \sum_{<i,j>} V(\mathbf{{\Omega}}_{i},\mathbf{{\Omega}}_{j})$,
where $V(\mathbf{{\Omega}}_{i},\mathbf{{\Omega}}_{j})$ is the orientational interaction between a pair of molecules (or molecular complexes in general) and $<i,j>$ denotes the nearest-neighbor molecules $i$ and $j$. Our calculations are based on the simplest attractive, ${\cal O}(3)$-invariant interaction
$V(\mathbf{{\Omega}}_{i},\mathbf{{\Omega}}_{j})$
that involves second-rank quadrupolar tensor with uniaxial ($\mathbf{\hat{T}}_0^{(2)}$) and biaxial ($\mathbf{\hat{T}}_2^{(2)}$) components, and a third-rank octupolar tensor ($\mathbf{\hat{T}}_2^{(3)}$).  The tensors are built out of
the  orthonormal tripod $\mathbf{{\Omega}}_{k}$ of vectors $\{ \mathbf{\hat{a}}_k, \mathbf{\hat{b}}_k, \mathbf{\hat{c}}_k \}$ defining the orientational degrees of freedom of the {\emph{k-th}} molecule. Under these assumptions
  $V(\mathbf{{\Omega}}_{i},\mathbf{{\Omega}}_{j})$ takes the form
\begin{eqnarray} \label{Hint}
V(\mathbf{{\Omega}}_{i},\mathbf{{\Omega}}_{j})=-\epsilon \left[ \left( \mathbf{\hat{T}}_0^{(2)}(\mathbf{{\Omega}}_{i}) + \sqrt2 \lambda \mathbf{\hat{T}}_2^{(2)}(\mathbf{{\Omega}}_{i}) \right) \cdot \right. \hspace{1cm} \nonumber \\
\left.\left( \mathbf{\hat{T}}_0^{(2)}(\mathbf{{\Omega}}_{j}) + \sqrt2 \lambda \mathbf{\hat{T}}_2^{(2)}(\mathbf{{\Omega}}_{j}) \right) +\,\,\, \tau \mathbf{\hat{T}}_2^{(3)}(\mathbf{{\Omega}}_{i})\cdot \mathbf{\hat{T}}_2^{(3)}(\mathbf{{\Omega}}_{j})
\right],
\end{eqnarray}
where \cite{Longa}
\begin{eqnarray} \label{tensors}
\mathbf{\hat{T}}_0^{(2)}(\mathbf{{\Omega}}_{k}) &=&  \sqrt{\frac{3}{2}} \left( \mathbf{\hat{c}}_k \otimes \mathbf{\hat{c}}_k - \frac{1}{3} \mathbf{1} \right), \nonumber
\\
\mathbf{\hat{T}}_2^{(2)}(\mathbf{{\Omega}}_{k}) &=&  \frac{1}{\sqrt2} \left( \mathbf{\hat{a}}_k \otimes \mathbf{\hat{a}}_k - \mathbf{\hat{b}}_k \otimes \mathbf{\hat{b}}_k \right),
\\
\mathbf{\hat{T}}_2^{(3)}(\mathbf{{\Omega}}_{k}) &=& \frac{1}{\sqrt6} \sum_{(\mathbf{\hat{x}},\mathbf{\hat{y}},\mathbf{\hat{z}})\in
\mathcal{\pi}(\mathbf{\hat{a}}_k,\mathbf{\hat{b}}_k,\mathbf{\hat{c}}_k)} \mathbf{\hat{x}} \otimes \mathbf{\hat{y}} \otimes \mathbf{\hat{z}},  \nonumber
\end{eqnarray}
and where $ \mathbf{\hat{T}}_{m}^{(L)} \cdot \mathbf{\hat{T}}_{m'}^{(L)} =  \delta_{mm'}$ \cite{cross-couplig}.
The summation in $\mathbf{\hat{T}}_2^{(3)}$  runs over all permutations of $(\mathbf{\hat{a}}_k,\mathbf{\hat{b}}_k,\mathbf{\hat{c}}_k)$. The symbol $`\cdot `$
denotes  the scalar product formed by a full contraction of the Cartesian indices, and
$\mathbf{\hat{T}}_m^{(L)}(\mathbf{{\Omega}}_{i}) \cdot \mathbf{\hat{T}}_{m'}^{(L)}(\mathbf{{\Omega}}_{j})=
\Delta^{(L)}_{mm'}$ are linear combinations of Wigner rotation matrices \cite{Longa}.

The interaction (1) can be interpreted within point dispersion forces approximation. The quadrupolar tensor
$\mathbf{\hat{T}}_0^{(2)} + \sqrt2 \lambda \mathbf{\hat{T}}_2^{(2)}$ is then
proportional to the anisotropic part of the dielectric polarizability tensor
of a molecule while $\mathbf{\hat{T}}_2^{(3)}$ is the $T_d$-symmetric component of the
third-order polarizability tensor. The model parameters $\{\epsilon,\lambda,\tau\}$ are related to the molecular absorption frequencies.

Special cases of the model have already been studied.  For   $\tau=\lambda=0$ the potential in Eq. (1) reduces to the well-known Maier-Saupe or Lebwohl-Lasher  \cite{Lebwohl} potential, which describes a phase diagram with isotropic and uniaxial nematic phases connected by a first-order phase transition. Second is the one where  $\lambda \ne 0$ and $\tau=0$.  Potential (1) then reduces to the model proposed by Luckhurst \emph{et al.} \cite{Luckhurst}, which was extensively studied by Biscarini {\emph{et. al.}} \cite{Zannoni}. The model describes a phase diagram with
uniaxial nematic and biaxial nematic phases connected by the second-order phase  transition. The phases include: a prolate  uniaxial ($N_{U+}$) phase,  an oblate
uniaxial ($N_{U-}$) phase, a biaxial ($N_{B}$) phase, and an isotropic ($I$) phase.
A self-dual point for which
$\lambda=1/\sqrt{6}$ \cite{Longa,Zannoni} separates a phase in which the molecules are of   distorted prolate form  ($\lambda<1/\sqrt{6}$) from a phase in which the molecules are of
distorted oblate form ($\lambda>1/\sqrt{6}$). Bates and Luckhurst  have proposed a  simple relation between $\epsilon$, $\lambda$ and the opening angle for bent-core molecules based on segmental second rank interactions \cite{Bates}.

When only the term proportional to $\epsilon\,\tau$ is retained in Eq.~(1), the resulting model corresponds to a purely tetrahedratic coupling. It was
introduced by Fel \cite{Fel1} and studied via Mean Field (MF) and Monte Carlo (MC) simulations by Romano \cite{Romano}. For this model,  MF theory
predicts a second order phase transition from $I$ to $T$ phase and MC simulations indicate a weak first-order transition from  $I$ to $T$.

Combining quadrupolar and octupolar interactions in Eq.~(1), yields new possibilities
of which the most notable one is the spontaneous breaking of chiral symmetry. Such a symmetry breaking is already manifested in the ground state properties of the interaction of Eq.(1).
Indeed, to be consistent with Eq.(1), the average molecular configurations  of two bent-core
molecules in a chiral phase  must belong to one of two configurations of opposite chirality
($0<\delta < \pi/2$), as shown in Fig.~1. Due to the global ${\cal O}(3)$-invariance of $H$, the two configurations of different chirality are the source of two homo-chiral domains,
which are present with equal probability in the system's configuration space.
When $\tau=0$ ($\delta=0$) or $\lambda = 0$ ($\delta = \pi/2$) the molecular configurations are nonchiral and can be brought into coincidence by a rotation. In particular, these configurations produce  $N_U$, $N_B$, $T$ and $N_T$ phases. In order for both chiral configurations to be equivalent, the  degrees of freedom of the  \emph{k-th} molecule should involve a rotation $\mathbf{{\Omega}}_k$ and parity $p_k= \mathbf{\hat{a}}_k \cdot \left( \mathbf{\hat{b}}_k \times \mathbf{\hat{c}}_k \right)=\pm 1$, thereby reflecting an ${\cal O}(3)$ symmetry. Consequently, the free energy for the system composed of $N$ such molecules is given by  $\beta F= - \ln Z$, where
%
$Z = \prod_{k=1}^{N} \left( \frac{1}{2} \sum_{p_k=\pm1} \int \mathrm{d}\mathbf{\Omega}_k \right) \exp\left[ -{\beta}H \right]$, 
 and $\beta = 1/(k_BT )$ is the inverse temperature.

We will apply two methods to determine phase diagrams from $F$. The methods are the MF approximation and the Metropolis MC simulations with a crucial step being the
identification of the order parameters. The identification is achieved by expanding
the one-particle distribution function $P(\emph{p}, \mathbf{\Omega})$ in a series of symmetry adapted, real $\Delta$-functions:
\begin{eqnarray}
\label{opdfexpD}
P(\emph{p}, \mathbf{\Omega}) = \sum_{L, m, m',s} \frac{2L+1}{8\pi^2}\, \overline{p_s\,\Delta^{(L)}_{m m'}} \, p_s \Delta^{(L)}_{m m'}(\mathbf{\Omega}),
\end{eqnarray}
with
%
$ \overline{\mathbf{X}}  =\frac{1}{2} \sum_{\emph{p} = \pm1} \int \mathbf{X}(\emph{p}, \mathbf{\Omega}) P(\emph{p}, \mathbf{\Omega}) \mathrm{d}\mathbf{\Omega}$, $p_s=(1,p)$,
%
and
\begin{eqnarray}\label{orth}
    \frac{1}{2} \sum_{\emph{p} \pm1}\int  \mathrm{d}\mathbf{\Omega}\,
p_s\Delta^{(L)}_{mn}(\mathbf{\Omega})\,\,
  p_{s'}\Delta^{(L')}_{m'n'}(\mathbf{\Omega})  =\frac{8\pi^2}{(2L+1)}
  \delta_{LL'}\delta_{mm'}\delta_{nn'}\delta_{ss'}.
\end{eqnarray}

For the interaction of Eq.~(1), the allowed symmetry reduction is shown in Fig.~2
along with the primary order parameters that acquire non-zero averages when crossing  various phase transitions. In order to be consistent  with $L=3$ primary tetrahedratic order parameter, we list all remaining (secondary) molecular order parameters of rank $L\le3$. These are:\emph{(a)}
$\overline{\Delta^{(2)}_{0 2}}$ for $N_U$;
\emph{(b)}
$\overline{\Delta^{(2)}_{2 0}}$
and
$\overline{\emph{p}\Delta^{(3)}_{2 2}}$ for  $N_B$;
\emph{(c)}
$\overline{\emph{p}\Delta^{(2)}_{2 0}}$ and $\overline{\emph{p}\Delta^{(2)}_{2 2}}$, along with $N_U$ secondary 
order parameters,
for $N_T$; and finally
\emph{(d)}
for $N_T^*$ all aforementioned order parameters are nonzero, in addition to 
$ \overline{\emph{p}}$, $\overline{\emph{p}\Delta^{(2)}_{0 0}}$, and  $\overline{\emph{p}\Delta^{(2)}_{0 2}}$.
Here
\begin{eqnarray}
  \Delta^{(L)}_{m m'} = \hspace{-1mm}\left(
\frac{1}{\sqrt2}\right)^{\hspace{-1mm}2+\delta_{0m}+\delta_{0m'}}\hspace{-7mm}
\sum_{s,s'=\pm1}\hspace{-4mm}\left[\delta_{ss'}+(-1)^L\delta_{-ss'}\right]\hspace{-1mm} {\cal{D}}_{sm s'm'}^L,
\end{eqnarray}
where ${\cal{D}}$ are the Wigner rotation matrices.

We are now in position to evaluate the MF approximation to $F$ and obtain the equilibrium properties of our system. We have identified
six stable phases of the model (1) that cover the whole
symmetry reduction flowchart of Fig.~2. The phases are: (a) the isotropic phase; (b) the uniaxial prolate or oblate
nematic phase;  (c) the biaxial phase; (d) the tetrahedratic phase;
(e) the  prolate or oblate
 tetrahedratic nematic  phase and
(f) the chiral tetrahedratic nematic phase.
Exemplary phase diagrams for $\tau=1$ and and for $\tau=9$  are shown in Figs.~3 and 4, respectively. Fig.~3 shows a phase diagram in which the high-temperature region is dominated by the nematic phases. Fig.~4 shows a phase diagram
in which occurs a direct $I-T$ phase transition.
Phase diagrams for intermediate values of $\tau$ can partly be deduced by extrapolation.
One of them is of particular interest. Namely, for $\tau=\frac{28}{15}$ six phases: $I$, $T$, $N_{T+}$, $N_{T*}$, $N_{T-}$ and $N_{U-}$ meet at a single multicritical Landau point.
 All phase transitions found are of second order except when more than
one order parameter acquires a nonzero value at a bifurcation.
Such first order phase transitions occur at $I-N_{T}$,  $T-N_{T}$ or  $I-N_{U}$.
A fuller account of the properties of this model is deferred to our future publication.

To check the validity of the MF predictions the phase diagrams were also determined from MC simulations. The simulations were performed
on a 16x16x16 lattice with periodic boundary conditions. Each MC move included a  rotation of a molecule's orientation and a  parity inversion.
The size of MC rotational moves was selected to produce an acceptance
ratio between $30\%$ and $40\%$ in the ordered phases. Typically,  50 000 to 200 000  lattice sweeps were used to thermalize the system and 60 000 to 200 000 sweeps for measurements. In the MC simulations, phase transitions were detected by observing the temperature   dependence of the order parameters. In turn, these quantities were  determined from the asymptotic behaviors of  the correlation functions
$G^{(L)}_{mm'} (|i-j|)=\overline{\mathbf{\hat{T}}_{m}^{(L)}(\mathbf{{\Omega}}_{i}) \cdot \mathbf{\hat{T}}_{m'}^{(L)}(\mathbf{{\Omega}}_{j})}$
and
$ G_{pp}(|i-j|)= \overline{p_i p_j}$ for large $|i-j|$.
Here the overline  indicates an ensemble average.
In the simulations, the temperature resolution $\delta t$ satisfied $\delta t=0.01$. At the above resolution  (and system's size) our simulations were unable to distinguish between weakly first-order and second-order phase transitions.
The MC transition curves lie below those obtained from MF. The  MF results worsen from a discrepancy of 8\% to about  20\% as $\lambda$ increases from $\lambda=0$  ($I-N_U$)
 to  $\lambda=0.3$ ($N_T-N_T^*$).
The $I-N_U$ part of our MC diagram  agrees with MC results of Biscarini {\emph{et al.} \cite{Zannoni}.

The new liquid crystal phases appear due to the combination of octupolar  interactions with quadrupolar contributions. In addition to a tetrahedratic nematic phase, existing in prolate ($N_{T+}$) and oblate ($N_{T-}$) versions, the  $N_{T}^{*}$ phase appears stable.  It does not appear  for $\lambda=0$ or $\lambda=\sqrt{\frac{3}{2}}$, which correspond to a uniaxial limit. From the leading MF contribution to the parity order parameter, $\overline{\emph{p}}$, an induced homogeneous chirality is seen to emerge when both  biaxial and tetrahedratic orders  condense. The estimation gives
\begin{equation}
\overline{\emph{p}} \cong \frac{\sqrt2}{210 t^4} \tau \lambda \left(-3 + 2 \lambda^2 \right) \overline{\Delta^{(3)}_{2 2}} \left( \overline{\Delta^{(2)}_{2 0}}+\sqrt2 \lambda \overline{\Delta^{(2)}_{2 2}} \right)^3 ,
\end{equation}
which vanishes when biaxial or tetrahedratic order parameters are zero. A maximal transition temperature to a chiral phase is achieved for the self-dual point of $\lambda=1/\sqrt{6}$.

To conclude, bent-core liquid crystals  can stabilize
the elusive thermotropic biaxial nematic phase \cite{Madsen,Acharya}, and reveal a path  to a series of new spatially homogeneous but anisotropic liquids \cite{Lubensky}. Of these liquids the most interesting one  is the $N_T^*$ phase, which should emerge from a nonchiral liquid as a result of spontaneous chiral symmetry breaking.  The simplest 'spin-spin'-type of  molecular model of Eq.~(1), with quadrupolar and octupolar interactions  supports this scenario, but,  unlike chiral phases of  ordinary chiral materials, the  orientational order in $N_T^*$ appears  spatially uniform. It is worth noting, however, that inclusion of
higher-order cross-coupling terms between uniaxial, biaxial and tetrahedratic interactions
can superimpose a spatial  modulation to  $N_T$ and $N_T^*$ \cite{cross-couplig}. Thus the structures stabilized in our model can serve as the long-wavelength limits to a family of spatially modulated chiral structures  that can possibly condense in the presence of higher-order interactions.  Owing to general form of the interaction, Eq.~(1), these conclusions should apply to any system where tetrahedratic and quadrupolar order may simultaneously coexist.
\begin{acknowledgments}
This work was supported by Grant N202 169 31/3455 of the Polish Ministry of Science and Higher Education, and by the EC Marie Curie Actions 'Transfer of Knowledge', project COCOS (contract MTKD-CT-2004-517186).
\end{acknowledgments}

\newpage

%

\begin{figure}[ht]\label{chiralSymmetry}
\begin{picture}(200,300)
  \put(-20,-10){
   \includegraphics{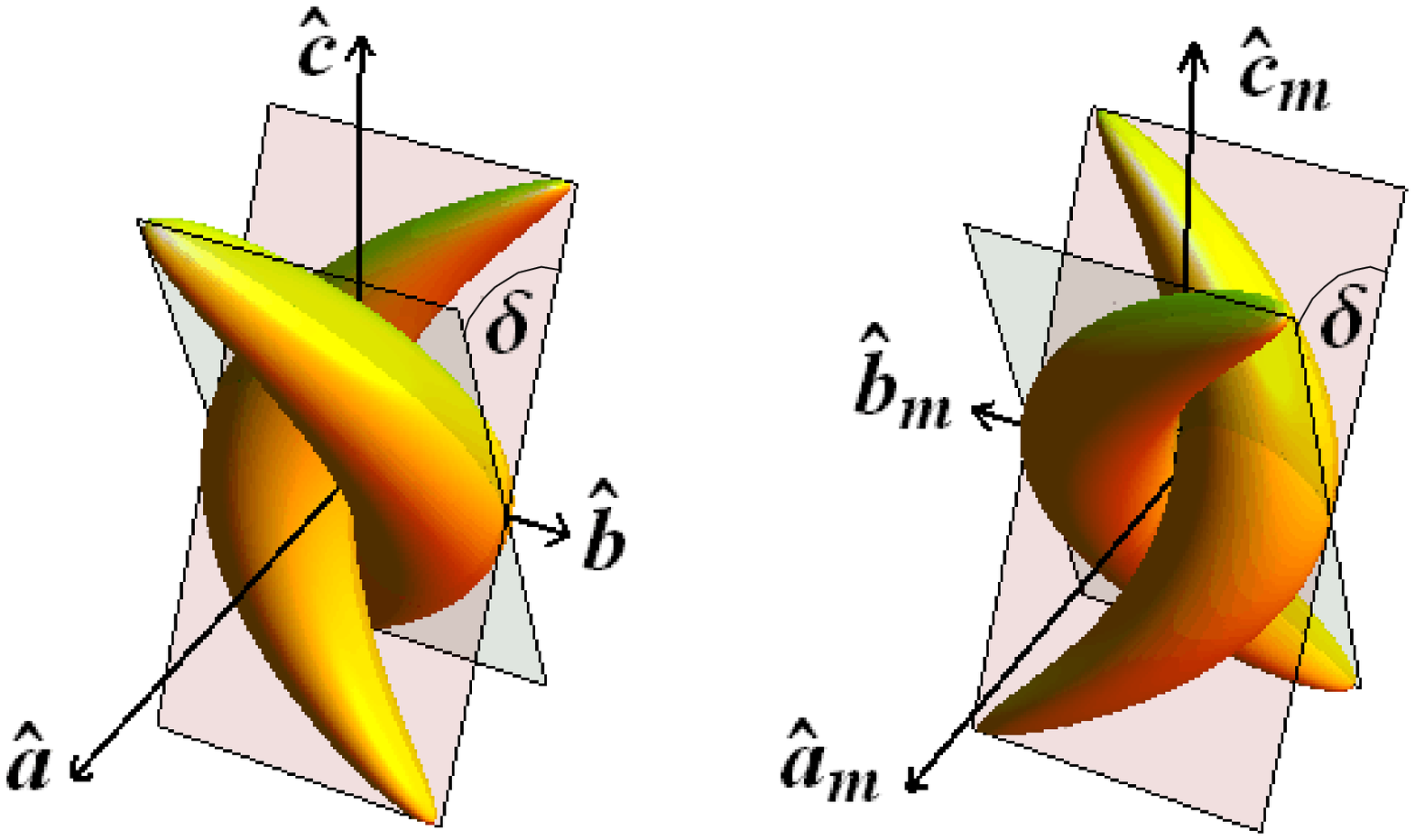}
  }
\end{picture}
\caption[]{(Color online) Ground state configurations of opposite chirality ($0<\delta < \pi/2$) for a pair of bent-core molecules  in $N_T^*$ phase;
the configurations for $0<\delta < \pi/2$ cannot be brought into coincidence by a proper rotation. $(\mathbf{\hat{a}}_m,\mathbf{\hat{b}}_m,\mathbf{\hat{c}}_m)$
are mirror images of $(\mathbf{\hat{a}},\mathbf{\hat{b}},\mathbf{\hat{c}})$.
}
\label{fTd}
\end{figure}


%
\begin{figure}[ht]
\begin{picture}(200,300)
  \put(-30,-10){
   \includegraphics{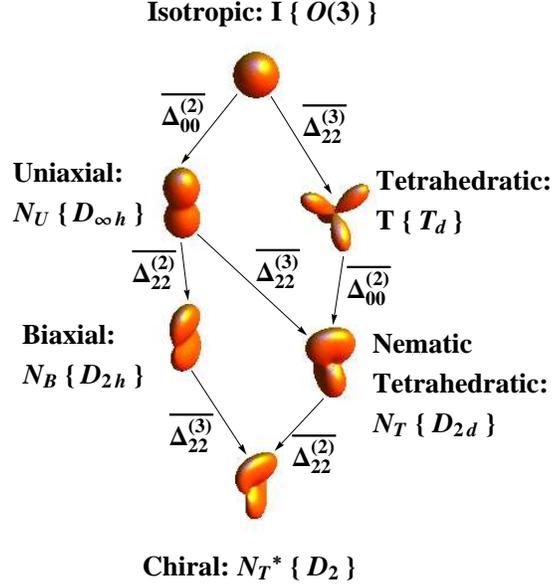}
  }
\end{picture}
\caption[]{ (Color online) A flowchart of phase transitions between liquid-crystal phases for  molecules
interacting through pair potential, Eq.~(1).
Primary order parameters, which become nonzero
at the transitions, their symmetry groups and abbreviated notation for the structures are indicated. In addition, sketched are one-particle spherical distribution functions $P\sim const + a Y_{20}+b(Y_{22}+Y_{2-2}) + c{\,{i}}(Y_{32}-Y_{3-2})$ illustrating symmetries considered.}
\label{fD2d}
\end{figure}

%

%
%
\begin{figure}[ht]
    \begin{picture}(10,300)
    \put(-140, -10){
      \includegraphics{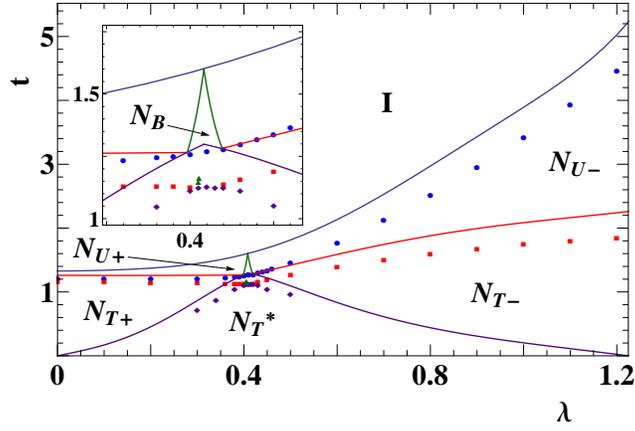}
    }
    \end{picture}
\caption[]{(Color online) Phase diagram for $\tau =1$. Lines represent MF results; points are from MC simulations for the three dimensional cubic lattice $(16 \times 16 \times 16)$;  $t=(\beta \epsilon)^{-1}$.} \label{Diagram1}
\end{figure}
%
%

%

\begin{figure}[ht]
\begin{picture}(300,300)
  \put(-20,-10){
   \includegraphics{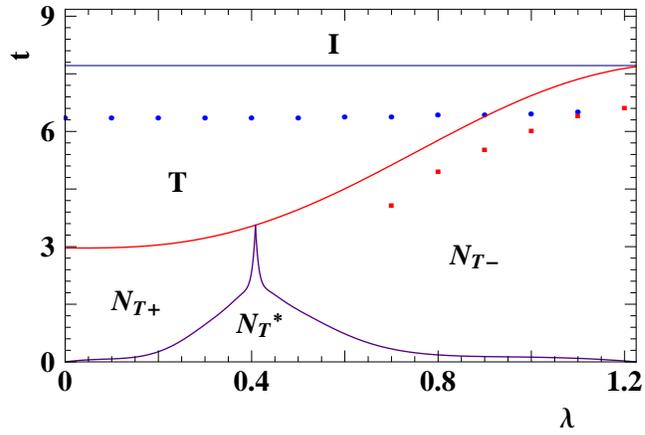}
  }
\end{picture}
\caption[]{(Color online) MF phase diagram for $\tau=9$ (see captions to Fig.~3). Simulations here are difficult since parity degrees of freedom condense to a glassy state. The $N_{T}^{*}$ phase cannot be reached by standard Metropolis simulations. }
\label{Diagram2}
\end{figure}


\begin{thebibliography}{999}
\bibitem{cross-couplig}
  We have left out interactions that involve products of more than two $L=2,3$ tensors, or couple to lattice directions.
\end{thebibliography}
\end{document}